# Structure, magnetism and magnetic compensation behavior of $Co_{50-x}Mn_{25}Ga_{25+x}$ and $Co_{50-x}Mn_{25+x}Ga_{25}$ Heusler alloys

G. J. Li (李贵江), E. K. Liu (刘恩克), Y. J. Zhang (张玉洁), Y. Du (杜音), H. W. Zhang (张宏伟), W. H. Wang*(王文洪), and G. H. Wu (吴光恒)

*Beijing National Laboratory for Condensed Matter Physics, Institute of Physics, Chinese Academy of Sciences, Beijing 100190, People's Republic of China*

Abstract：

The structure, magnetism, magnetic compensation behavior, exchange interaction and electronic structures of $Co_{50-x}Mn_{25}Ga_{25+x}$ and $Co_{50-x}Mn_{25+x}Ga_{25}$ ($x$=0-25) alloys have been systematically investigated by both experiments and first-principles calculations. We found that all the samples exhibited body centered cubic structures with a high degree of atomic ordering. With increasing Ga content, the composition dependence of lattice parameters shows a kink point at the middle composition in $Co_{50-x}Mn_{25}Ga_{25+x}$ alloys, which can be attributed to the enhanced covalent effect between the Ga and the transition metals. Furthermore, a complicated magnetic competition has been revealed in $Co_{50-x}Mn_{25}Ga_{25+x}$ alloys, which causes the Curie temperature dramatically decrease and results in a magnetic moment compensation behavior. In $Co_{50-x}Mn_{25+x}Ga_{25}$ alloys, however, with increasing Mn content, an additional ferrimagnetic configuration was established in the native ferromagnetic matrix, which causes the molecular moment monotonously decreases and the exchange interaction enhances gradually. The electronic structure calculations indicate that the $Co_{50-x}Mn_{25+x}Ga_{25}$ alloys are likely to be in a coexistence state of the itinerant and localized magnetism. Our study will be helpful to understand the nature of magnetic ordering as well as to tune magnetic compensation and electronic properties of Heusler alloys.

# I. INTRODUCTION

Full-Heusler ternary compounds of $X_2YZ$ composition (with X, Y the transition metals and Z the main-group element) have been well studied for decades with very rich physical properties, ranging from shape memory effect[1-6], over half-metallicity[7,8], topological insulators[9] to thermoelectric materials[10] and superconductivity[11,12]. The crystal lattice of Heusler compounds is described by the space group $Fm\bar{3}m$, with the atomic configuration shown in Fig. 1. The unit cell is composed of four interpenetrating face centered cubic (F.C.C) sublattices *A*, *B*, *C*, *D*, whose Wyckoff coordinates are (0 0 0), (1/4 1/4 1/4), (1/2 1/2 1/2) and (3/4 3/4 3/4), respectively. When X atoms have relatively more valence electrons than Y atoms in Heusler alloy, they preferentially occupy the two equivalent crystallographic sites *A* and *C* while Y and Z atoms occupy *B* and *D* sites, respectively. The alloy constructs a $L2_1$ structure, i.e. $Cu_2MnAl$-type structure. When the valence-electron number of X atoms is less than that of Y atoms, however, the alloy crystallizes in a $Hg_2CuTi$-type structure, usually considered as a variant of $L2_1$ structure, in which X atoms occupy *A* and *B* sites and Y atoms occupy *C* sites [13]. This phenomenon has been considered as an empirical rule for atomic preferential occupation in Heusler alloys and attributed to the covalent effect caused by the *p-d* orbital hybridization between the main-group and the transition-metal atoms[14-17]. The validity of this rule has been also confirmed by the minimum of calculated total energy for the electronic structure of Heusler alloys with different atomic configurations[15,18].

In general, Co, Mn, and Ga can compose three kinds of stoichiometric Heusler alloys, $Co_2MnGa$, $CoMnGa_2$ and $Mn_2CoGa$. Due to the different valence electron concentration, e/a, they can crystallize in two different structures: the $Cu_2MnAl$-type for the former two and $Hg_2CuTi$-type for the latter one[19-21]. The physical properties, such as magnetism and high spin polarization of these alloys with the stoichiometric composition, have been investigated in detail[20,22]. However, there are no systematically experimental and theoretical studies on the off-stoichiometric alloys from $Co_2MnGa$ to $CoMnGa_2$ and/or to $Mn_2CoGa$.

The Heusler structure has the interesting feature of forcing the Mn atoms to be the first and/or second nearest neighbors of each other. This allows them to couple in a ferromagnetic (FM)/ferrimagnetic (FIM) manner instead of the normal antiferromagnetic (AFM) order. Moreover, due to the different atomic configuration and the tunable covalent effect, the rich physical phenomena related to the structural and magnetic properties can be expected in the off-stoichiometric alloys with continuously changing composition: 1) The structural characters during the structure transition between the $Cu_2MnAl$-type and the $Hg_2CuTi$-type based on the atomic preferential occupation rule as mentioned above; 2) the microscopic understanding of the nature of the transition from the FM $Co_2MnGa$ to the FIM $Mn_2CoGa$ and $CoMnGa_2$; 3) the role conversion of magnetic contributor and the resulted magnetic compensation behavior between $Co_2MnGa$ and $Mn_2CoGa$, $CoMnGa_2$; 4) the changes of exchange interaction during the variation of the atomic configuration, especially when the local FIM structure was built in a FM matrix; 5) the change of electronic and magnetic structures due to the $d$-$d$ and $p$-$d$ hybridization competition. These facts motivated us to investigate the series of off-stoichiometric samples of $Co_{50-x}Mn_{25}Ga_{25+x}$ and $Co_{50-x}Mn_{25+x}Ga_{25}$ ($x$=0-25) by both experiments and first-principles calculations. In this paper we present a detailed study of the crystal structure, magnetism and magnetic compensation behavior of $Co_{50-x}Mn_{25}Ga_{25+x}$ and $Co_{50-x}Mn_{25+x}Ga_{25}$ ($x$=0-25) alloys. Based on the preferential atomic occupations and the highly degree of ordering of the two series alloys, the relationship between the magnetic exchange interaction and electronic structures are also discussed.

The paper is organized as follows: Section II contains details concerning the methods of the experiment and the calculations. Section III presents the experimental and calculated results, including: (1) the crystal structure, degree of ordering and atomic configurations; (2) the molecular and atomic magnetic moment, and magnetic structure; (3) the magnetic exchange interaction; (4) calculated electronic structure. Section IV presents a detailed discussion about the tunable magnetic moment compensation behaviour and possible compensated-ferrimagnetic half-metallicity. Finally, the paper is summarized in Sec. V.

## II. EXPERIMENTAL AND CALCULATION DETAILS

The two series of samples of $Co_{50-x}Mn_{25}Ga_{25+x}$ and $Co_{50-x}Mn_{25+x}Ga_{25}$ ($x$ =0-25) were selected as the studied objects which connect three stoichiometric Heusler alloys of $Co_2MnGa$, $Mn_2CoGa$ and $CoMnGa_2$. These alloys were synthesized by arc-melting elementary metals of Mn, Co and Ga with purity of 99.99% or higher under argon atmosphere. The as-cast ingots were sealed in evacuated quartz tubes with argon atmosphere. The samples were annealed at 1073 K for two days for homogenizing and then at 973K for one day for atomic ordering. Finally, the samples were quenched into ice water. The order-disorder phase transition temperatures were determined by the differential scanning calorimetry (DSC). The X-ray diffraction (XRD) examination with Cu-$K_\alpha$ was carried out with step-scan method to characterize the crystal structure and determine the lattice parameters. The magnetic properties were measured by a superconducting quantum interference device (SQUID) magnetometer and a vibrating sample magnetometer (VSM). The theoretical molecular magnetic moments and electronic structures were calculated using the Korringa-Kohn-Rostoker coherent-potential approximation based on the local density approximation (KKR-CPA-LDA)[23, 24].

## III. RESULTS

### A. Order-disorder transition of $Co_{50-x}Mn_{25}Ga_{25+x}$ and $Co_{50-x}Mn_{25+x}Ga_{25}$ alloys

It is very important to confirm the order-disorder transition temperature of our two series samples in order to determine the optimal annealing temperature for atomic ordering. In Fig.2 (a) and (b), we show the DSC curves of $Co_{50-x}Mn_{25}Ga_{25+x}$ and $Co_{50-x}Mn_{25+x}Ga_{25}$ alloys, respectively. For $Co_2MnGa$ alloy, the order-to-disorder transition temperature, i.e., from $L2_1$ to $B2$ occurs at about 1180 K, which is consistent with previous reports[25-27]. With the substitution of Mn or Ga for Co in the two serials of alloys, the transitions from $L2_1$ to $B2$ or $Hg_2CuTi$ to $B2$ all occur at temperatures higher than 1000 K. For the alloy of $x$=7.5 in $Co_{50-x}Mn_{25}Ga_{25+x}$, with increasing temperature, no transition from $L2_1$ to $B2$ was clearly observed in the DSC curves up to the melting temperature ($T_m$), which may suggest the high-ordered structure of these compounds stable up to the $T_m$ [26]. The atomic ordering temperature was then chosen at 973 K, which is lower than the $L2_1$ to $B2$ or $Hg_2CuTi$ to $B2$ transition temperature. Thus, all the samples quenched from this temperature should have a high-ordered $L2_1$ or $Hg_2CuTi$-type structure. In the following section, we will quantitatively discuss the degree of atomic ordering in these samples based on the XRD intensities.

### B. High-ordered structure in $Co_{50-x}Mn_{25}Ga_{25+x}$ and $Co_{50-x}Mn_{25+x}Ga_{25}$ alloys

In principle, the atomic occupation in Heusler alloys is determined by its valence-electron number[14-16]. In the present work, the sequence of the valence-electron number is: (Co) $3d^74s^2$ >(Mn) $3d^54s^2$ > (Ga)$4s^24p^1$. Following the atomic preferential occupation rule, the Co$_2$MnGa alloy crystallizes in $L2_1$ structure in which the atomic sequence in the nearest neighbor (denoted as 1nn) distance along the [111] direction is Co(*A*) Mn(*B*) Co(*C*) Ga (*D*). When Co atoms were substituted by Mn atoms to obtain the alloys of Co$_{50-x}$Mn$_{25+x}$Ga$_{25}$, the newly introduced Mn atoms trend to occupy the Co (*A*) sites. Thus, the final substitution will produce an alloy of Mn$_2$CoGa. It forms the Hg$_2$CuTi-type structure with the atomic sequence of Mn(*A*) Mn(*B*) Co(*C*) Ga(*D*) in [111] direction[20]. For the case of the substitution of Ga for Co atoms in Co$_{50-x}$Mn$_{25}$Ga$_{25+x}$ alloys, the newly introduced Ga atoms have less valence electrons than the origin Mn atoms. Consequently, the new Ga atoms will occupy Mn (*B*) site, which forces the original Mn(*B*) atoms to occupy the vacant Co(*A*) sites. This creates an atomic sequence of Mn(*A*) Ga(*B*) Co(*C*) Ga(*D*) (equal to Ga(*A*) Mn(*B*) Ga(*C*) Co(*D*)).Consequently the alloy finally becomes CoMnGa$_2$ with the Cu$_2$MnAl-type structure.

In order to prove the atomic preferential occupation mentioned above, the theta-2theta X-ray diffraction (XRD) patterns of Co$_{50-x}$Mn$_{25}$Ga$_{25+x}$ and Co$_{50-x}$Mn$_{25+x}$Ga$_{25}$ alloys measured at room temperature are shown in Fig. 3. The XRD patterns confirm that all samples are in single phase with the body centered cubic (B.C.C) structure. Besides the principal reflection peaks of (220), (400), (422) and (440), the superlattice reflection peaks of (111) and (200) can be also observed clearly, which indicates that high-ordered structure is obtained in all samples (see supplementary Figs.SI and SII). Especially for CoMnGa$_2$, it also shows the superlattice reflection peaks of (111), (200), (311), (222) and (331), as marked in Fig.2 (a). This indicates that the CoMnGa$_2$ alloy forms a high $L2_1$–ordered structure[21].

Combined with experimental and simulated XRD results, the degree of ordering S for $L2_1$ or

Hg$_2$CuTi can be roughly estimated by following function[28]:

$$S^2 = \frac{I_{exp.}(111)/I_{exp.}(220)}{I_{cal.}(111)/I_{cal.}(220)}, \quad (1)$$

in which I (111)$_{exp}$/ I (111)$_{cal.}$ and I (220)$_{exp}$/I (220)$_{cal.}$ are, experimental and calculated, integrated intensity of superlattice peaks and fundamental peaks, respectively. The results were shown in the insets of Fig.3. As we excepted, in the two serials of alloys, the values of ordering parameter S are all higher than 0.9, indicating a highly ordered $L2_1$ or Hg$_2$CuTi-type structure.

Additionally, we can find that when the composition changes from CoMnGa$_2$ to Co$_2$MnGa, the intensity of (311), (222), and (331) superlattice reflection peaks becomes weak, whereas the reflections peaks of (111) and (200) still can be observed clearly. Compared with Co$_{50-x}$Mn$_{25}$Ga$_{25+x}$ samples, the Co$_{50-x}$Mn$_{25+x}$Ga$_{25}$ samples show relatively weak superlattice reflection peaks, as shown in Fig.3 (b). This is due to that the scattering factors of Co and Mn atoms are closed to each other[29, 30].

Indexing the XRD patterns, we obtained the lattice parameters of all samples and listed them in Table I. The composition dependence of lattice parameters of the two series of samples is shown in Fig. 4. One can find that the lattice parameters of the two series of samples increase monotonously with the substitution of Ga or Mn for Co atoms. This is mainly due to that the atomic radii of Ga (0.181nm) and Mn (0.179nm) are larger than that of Co atom (0.167nm)[31]. Linearly fitting the experimental data, we found that the slope of the curve is a constant for Co$_{50-x}$Mn$_{25+x}$Ga$_{25}$ alloys. However, the slope suddenly decreases for Co$_{50-x}$Mn$_{25}$Ga$_{25+x}$ alloys at the middle composition of $x$ = 12.5, showing a kink behavior. In our pervious work[32], an unusual lattice parameter changes has been also found in Mn$_{50-x}$Co$_{25}$Ga$_{25+x}$ Heusler alloys by substituting Ga for Mn. As we will discuss below, the decrease of the lattice constant originates from the enhanced covalent effect between transition-metal and main-group atoms in the Heusler system.

In fact, as shown in Fig. 1, Heusler alloy can be also considered as a close-packed structure constructed from the *A/C* and *B/D* atomic layers with a distance of one fourth of the lattice parameter. From this point of view, $Co_2MnGa$ is composed of Co/Co and Mn/Ga layers and the $CoMnGa_2$ consists of Ga/Ga and Co/Mn layers in $Cu_2MnAl$-type structure, while $Mn_2CoGa$ has Co/Mn and Mn/Ga layers in $Hg_2CuTi$-type structure. For $Co_{50-x}Mn_{25}Ga_{25+x}$ alloys, the kink behavior corresponds to a structural mutation occurred at the middle composition: a Ga/Ga layer has been created on one side of the Co/Mn layer and a transition from Ga/Mn layer to Ga/Ga layer on another side will start when we further increase Ga content. Thus, the abnormal increment of lattice parameter can be attributed to the *p-d* orbital hybridization between *p*-block elements (here is Ga) and transition metals[17,32]. The covalent bonding effect enhanced by the appearance of the Ga/Ga atomic layers in $Co_{50-x}Mn_{25}Ga_{25+x}$ should be stronger than that in $Co_{50-x}Mn_{25+x}Ga_{25}$ alloys. It is because, for $Co_{50-x}Mn_{25+x}Ga_{25}$ alloys, only the transition from Co/Co layers to Co/Mn layers occurs and the Mn/Ga layers keep unchanged.

**C. Molecular and atomic magnetic moment**

Figures 5 (a) and (b) show the representative isothermal magnetization curves of $Co_{50-x}Mn_{25}Ga_{25+x}$ and $Co_{50-x}Mn_{25+x}Ga_{25}$ alloys, respectively. We found that ,when $x \geqslant 17.5$ in $Co_{50-x}Mn_{25}Ga_{25+x}$ alloys, the magnetization values of these alloys linearly increase with the increasing of external magnetic field due to the contribution from conduction electrons polarization. This magnetization behavior is obviously different from that in other alloys. By extrapolating the linear part of M(H) curves to zero field, the magnetization valves under zero field were obtained and these values can be considered as the saturation magnetization value.

Collecting these magnetization values, the composition dependence of the molecular magnetic moments of $Co_{50-x}Mn_{25}Ga_{25+x}$ and $Co_{50-x}Mn_{25+x}Ga_{25}$ alloys were obtained and shown in Fig.5(c). The theoretical data for $Co_{50-x}Mn_{25+x}Ga_{25}$ and $Co_{50-x}Mn_{25}Ga_{25+x}$ alloys were obtained by first principle calculations based on KKR-CPA-LDA method and shown in Table II and Table III, respectively. The experimental values of $Co_2MnGa$ and $Mn_2CoGa$ are consistent with those reported before[20, 33, 34]. The experimental and calculated results indicate that the Heusler alloy $CoMnGa_2$ is in a FIM structure, in which the moment of Mn atom antiparallelly aligns with that of Co atom. Therefore, the competition between antiparallel moments of the two sublattices gives rise to a net molecular magnetic moment of 1.34$\mu_B$. By substitution of Mn or Ga for Co atoms in $Co_{50-x}Mn_{25+x}Ga_{25}$ and $Co_{50-x}Mn_{25}Ga_{25+x}$ alloys, different atomic occupation will bring out various values of molecular magnetic moment.

In $Co_{50-x}Mn_{25+x}Ga_{25}$ alloys, the molecular magnetic moment decreases monotonously with the increase of Mn content. This is due to that the newly introduced Mn atoms will replace the Co ($A$) atoms and occupy the $A$ sites, at which the magnetic moments of Mn($A$) atoms antiparallelly align with those of the Mn ($B$) and Co($C$) atoms, as shown in Fig.6. Thus, a local FIM structure is added to the native FM matrix, just like the case of $Ni_{50-x}Mn_{25+x}Ga_{25}$ alloys[35], which causes the molecular magnetic moment of the alloys monotonously decrease. As a result, with increasing Mn contents, the $Co_{50-x}Mn_{25+x}Ga_{25}$ alloys will change from ferromagnetic $Co_2MnGa$ to compensated ferrimagnetic[36] $Mn_2CoGa$ due to the competition between the two antiparallel-aligned sublattices of Mn($A$) and the native Co($A$,$C$) /Mn($B$).

Very interestingly, the variation of molecular magnetic moments of $Co_{50-x}Mn_{25}Ga_{25+x}$ alloys is not monotonous, showing a minimum at $x$=17.5. This is due to that $Co_2MnGa$ and $CoMnGa_2$ are in the FM and the FIM structures, respectively, as shown in Fig.6. For $Co_2MnGa$, the parallelly-aligned magnetic moments of Co ($A$), Co ($C$) and Mn ($B$) atoms contribute to the total molecular magnetic moment. However, for $CoMnGa_2$, the main magnetic contributor is Mn($A$) atom whose moment is partly counteracted by antiparallelly-aligned magnetic moment of Co($C$) atom. It means that the FIM structure in $CoMnGa_2$ takes an "AFM" role in comparison with the FM $Co_2MnGa$ alloy during the chemical composition changes from $Co_2MnGa$ to $CoMnGa_2$. The minimum of the molecular magnetic moments at $x = 17.5$ indicates the magnetic moments of the two sublattices almost completely counteract each other, showing a magnetic compensation behavior. The unusual increase of molecular magnetic moment at $x > 17.5$ can therefore be ascribed to conversion of dominating role between the FM and the FIM structures.

In Table II and Table III, we also list the calculated atomic magnetic moments of the two series of alloys, respectively. Based on these data, we can conclude: 1) In the two series of alloys, Mn($A$) and Mn($B$) atoms carry different magnetic moments of about $1.9\mu_B$ and $2.8\mu_B$, respectively; 2) The magnetic moments of the Mn($A$) and Mn($B$) atoms possess the opposite signs, in other words, their magnetic moments are antiparallelly aligned;  3) The Co atom has a smaller magnetic moment ($<1$ $\mu_B$) than Mn atoms; 4) In both $Co_{50-x}Mn_{25+x}Ga_{25}$ and $Co_{50-x}Mn_{25}Ga_{25+x}$ alloys, the magnetic moments of the Mn($B$) atoms are quite consistent with each other, which can be attributed to their similar magnetic environment in the nearest neighboring distance where there is relatively low level of $p$-$d$ hybridization between Mn($B$) and Ga atoms. On the other hand, the moments of Mn($A$) show a relatively large variation in both series of alloys, especially in the $Co_{50-x}Mn_{25}Ga_{25+x}$ alloys where a complete covalent bonding finally forms between Mn($A$) and Ga atom in 1nn distance of Mn($A$)[37], as mentioned in the former section; 5) In $Co_{50-x}Mn_{25+x}Ga_{25}$ alloys, the magnetic moments of Co atoms increase markedly during the system changes from the FM to the FIM structure. It may be attributed to the transition of the dominant exchange interaction contributor from the Co-Mn to Mn-Mn atom pairs; 6) In $Co_{50-x}Mn_{25}Ga_{25+x}$ alloys, the magnetic moments of Co($A$) and Co($C$) atoms show a dramatic decrease, especially the latter, because the 1nn environments of them change from magnetic to non-magnetic finally. Obviously, the changes of covalence play an important role on the atomic magnetic moment, as well as the molecular magnetic moment.

We should point out here that, the high consistence between the experimental and calculated molecular magnetic moments indicates all the samples studied in this work crystallized in high-ordered structures, suggesting the atomic preference occupation rule is valid. We therefore propose that magnetic measurement may be considered as another criterion for testing the degree of ordering in Heusler compounds[38].

### D. Magnetic exchange interaction

Figure 7 shows the composition dependence of Curie temperature ($T_C$) of $Co_{50-x}Mn_{25}Ga_{25+x}$ and $Co_{50-x}Mn_{25+x}Ga_{25}$ alloys. When the alloy changes from $Co_2MnGa$ to $Mn_2CoGa$, the $T_C$ shows a monotonous increase with a kink at $x$ = 12.5. On the contrary, when the alloy changes from $Co_2MnGa$ to $CoMnGa_2$, the $T_C$ shows a remarkable decrease from 677 K to 200 K with a minimum of 150 K at the composition $x$ = 20. These results indicate that the two series of alloys show very different exchange interaction behavior.

In view of the large spatial separation of the Mn atoms (>0.4 nm) in Heusler alloys, the electrons of the unfilld Mn 3$d$ shell can be considered as very localized, so that the 3$d$-electrons belonging to different Mn atoms do not overlap considerably. The ferromagnetism in Heusler alloys is thought to arise from an indirect interaction between the Mn moments by the way of conduction electrons[39-41]. However, this model is unable to explain the experiments for the exchange interaction of atoms in close distances (e. g., the nearest and next nearest neighbor). The work of Stearns and her coworkers[42-44], which takes into account the neighboring relationship between transition-metal atoms, can be used to discuss the exchange interaction in these off-stoichiometric alloys. Based on their investigations on the itinerant 3$d$-electrons ($d_i$) and the conduction electron polarization, the 1nn atom pair of Mn-Mn has stronger exchange interaction than 1nn Mn-Co atom pair, and the 2nn Co-Mn atom pair has the weakest exchange interaction in the alloys studied in the present work.

Based on the mean-field approximation, the $T_C$ can be described as following [25, 45, 46], $T_C^{calc} = \frac{2J_o}{3k_B}$, in which $J_o = \sum_{i \neq o} J_{io}$. Here $k_B$ is the Boltzmann constant, and $J_o$ is effective coupling constant, that is, the sum of exchange interaction of the system. Thus, the changes of $T_C$ in the two systems can be determined by the variation of $J_o$.

In $Co_{50-x}Mn_{25+x}Ga_{25}$ alloys, the newly introduced Mn atoms occupy $A$ sites. With the increase of Mn content, the long-range FIM coupling between Mn($A$) and Mn($B$) sublattices in 1nn distance is gradually established. Since the coupling strength of Mn-Mn is stronger than that of Co-Mn, the overall exchange interaction is enhanced. This is the reason why the $T_C$ increases from 677 K to 710 K in $Co_{50-x}Mn_{25+x}Ga_{25}$ alloys. Additionally, the kink point at the middle composition $x = 12.5$ indicates the diversification in role of the FM coupling between Mn($B$) and Co($A$, $C$), and the FIM coupling between Mn($A$) and Mn($B$), which dominates the exchange interaction.

The variation of the $T_C$ in $Co_{50-x}Mn_{25}Ga_{25+x}$ alloys is another case. Based on the discussion about the character of atomic configuration mentioned above, with increasing Ga content, the magnetic atom of Co is substituted by the nonmagnetic atom of Ga. Meanwhile, the Mn($B$) atoms will be forced to move to the $A$ sites. And thus an FIM coupling between Mn($A$) and Mn($B$),Co($C$) atoms is gradually established within the network of the native FM coupling of Co($A$)/Co($C$)-Mn($B$). Although the added FIM coupling is stronger than the original FM one, it is at the cost of decrease of the number of Co atoms and destroys the original FM structure. Therefore, with the substitution of Ga for Co, the competition between the two magnetic structures cause the $T_C$ drastically decrease. At the composition of $x = 20$, the $T_C$ shows a minimum of 150 K (see Fig. 7) corresponding to the complete magnetic compensation in magnetism, indicating a very weak exchange interaction. In this situation, the mentioned weak FIM coupling of Mn($A$)-Co($C$) in 2nn distance becomes dominant for the magnetic interaction and increases the subsequent $T_C$.

### E. Calculated electronic structure

In order to further understand the origin of the experimentally observed magnetic interaction, we calculate the density of states (DOS) of $Co_{50-x}Mn_{25+x}Ga_{25}$ and $Co_{50-x}Mn_{25}Ga_{25+x}$ alloys, and the results are shown in Fig. 8 and Fig. 9, respectively. It can be seen that the states ranging from -6 eV to +3 eV mainly consist of $d$-electrons of Co and Mn atoms. The strong hybridization between $d$-electrons of Co and Mn atoms results in the dispersed occupied states, and the small contribution from $p$-electrons of Ga atoms should be included. In $Co_2MnGa$ alloy, there is remarkable hybridization between spin-up $d$-electrons of Mn($B$) and Co($A,C$) atoms, showing two strong hybridized peaks around -0.35 and -2.75 eV in spin-up direction. Thus, $Co_2MnGa$ shows a high $T_C$. The DOS distribution for $d$-electrons of Mn($A$) atom shows a plateau in a wide range of energy, which indicates the itinerant magnetism of Mn($A$) atoms. These show different exchange interaction among magnetic atoms and the coexistence of itinerant and localized magnetism in $Co_{50-x}Mn_{25+x}Ga_{25}$ alloys[47, 48].

For $Co_{50-x}Mn_{25+x}Ga_{25}$ alloys, as shown in Fig. 8, the *p-d* hybridization strength between Mn(*A*) and Ga(*D*) atoms (in 1nn) is stronger than that between Mn(*B*) and Ga(*D*) atoms (in 2nn), which results in the higher spin polarization level of Mn(*B*) than that of Mn(*A*)[20, 49]. This is the physical mechanism why the atomic magnetic moment of Mn(*B*) is larger than that of Mn(*A*).

For $Co_2MnGa$ alloy, the Mn(*B*) atom locates in a cubic crystal field with eight Co atoms as its 1nn atoms. With the substitution of Mn for Co(*A*) atoms in $Co_{50-x}Mn_{25+x}Ga_{25}$ alloys, the symmetry of the cubic crystal field is gradually broken, which results in the reduction of degeneracy of the energy level. Therefore, the DOS distribution for the *d*-electrons of Mn(*B*) atom becomes much broader and is not so localized as in $Co_2MnGa$ alloy. Due to the high level of hybridization between *d*-electrons of Mn(*B*) and Co(*C*) through the whole composition range, the broader DOS distribution for the *d*-electrons of Mn(*B*) atom results in an enhancement of hybridization between *d*-electrons of Mn(*A*) and Mn(*B*) atoms. Thus, the overall exchange interaction among all magnetic atoms can be enhanced in $Co_{50-x}Mn_{25+x}Ga_{25}$ alloys, although the coexistence of FM and FIM structure appears in the system. This results in the enhancement of experimentally observed $T_C$ in $Co_{50-x}Mn_{25+x}Ga_{25}$ alloys.

In $Co_2MnGa$, Ga(*D*) atom locates in a cubic crystal field with eight Co atoms as its 1nn atoms. Chemically substituting Ga for Co atoms in $Co_{50-x}Mn_{25}Ga_{25+x}$ alloys，the symmetry of this crystal field is broken progressively. It decreases the degeneracy of the energy level. So the DOS distribution for the *p*-electrons of Ga(*D*) atom becomes much broader and shifts to the higher energy states. As shown in Fig.9(g), the peak intensity of the DOS for *p*-electrons of Ga(*D*) atom around -2.8 and -4.8 eV in spin-down direction and around -5 eV in spin-up direction decrease with the changes of the composition. Meanwhile, the DOS distribution for *p*-electrons of Ga atom shows a wide plateau in the higher energy state. This shows the effect of the number of transition-metal atoms on the level of hybridization between *d*-electrons of magnetic atoms and *p*-electrons of Ga atoms. Moreover, the DOS distribution for spin-up and spin-down *d*-electrons of Co(*A*) atoms become much broader while those of Mn(*A*) and Co(*C*) atoms become much more localized with the substitution of Ga for Co atoms, which indicates that covalent hybridization level between *p*-electrons of main-group Ga atoms and *d*-electrons of transition-metal atoms becomes higher. Just due to the higher *p-d* covalent hybridization level, the atomic magnetic moments of Co(*A*) and Co(*C*), especially the latter, rapidly decrease with the composition , as shown in Table III. In addition, there are distinct hybridized peaks near the Fermi level in the DOS for *d*-electrons of Mn(*A*) and Co(*C*) atoms, and *p*-electrons of Ga atoms in Ga-rich alloys, especially in CoMnGa$_2$ alloy, which means an instable state. The higher states near the Fermi level in Ga-rich alloys make the ferrimagnetism unstable and can promote a noncollinear AFM configuration to reduce the DOS at Fermi level[50], just as reported results in CoMnSi alloys[51]. This may be the reason why the magnetization value of Ga-rich alloys shows a linear increase even under the high magnetic field [see Fig. 5 (a)]. The reported spin glass in CoMnGa$_2$ alloy can correlate to the occurrence of AFM configuration[21].

## IV DISCUSSION

We now discuss the atomic magnetic moment of Co$_{50-x}$Mn$_{25+x}$Ga$_{25}$ and Co$_{50-x}$Mn$_{25}$Ga$_{25+x}$ alloys. As shown in Figs. 8 and 9, the bonding states and antibonding states of Co atom mainly localize below the Fermi level ($E_F$) in both spin-up and spin-down direction. Above the Fermi level, the density of antibonding states for Co atom is very weak and lower than the antibonding states of Mn($B$) in energy. It indicates the exchange splitting of Co atoms is very weak, which results in a small magnetic moment. For DOS of Mn($B$) in spin-up direction, both bonding and antibonding states are below the $E_F$. Due to the strong exchange splitting, the antibonding states in spin-down direction are shifted above the $E_F$. Therefore, the magnetic moment of Mn($B$) is very large. The DOS of Mn($A$) is just contrary to that of Mn($B$), indicating the moment of Mn($A$) antiparallelly aligns with that of Mn($B$). As a result, the moment of Mn($A$) is relatively large. The calculation results shown in section II are consistent with the discussion above. The atomic moments of Mn and Co in Co$_2$MnGa and Mn$_2$CoGa well agree with the previous reports[20, 33, 47]. <span style="color:red">Moreover, as we will discuss below, in Co$_{50-x}$Mn$_{25+x}$Ga$_{25}$ and Co$_{50-x}$Mn$_{25+x}$Ga$_{25+x}$ alloys, the magnetic compensation behavior indeed origins from the appearance of the FIM configuration in $A$ site.</span>

In Co$_{50-x}$Mn$_{25}$Ga$_{25+x}$ alloys, the magnetic moment compensation and magnetic exchange interaction compensation were simultaneously observed in Ga-rich alloys. The occurrence of magnetic compensation behavior indicates the conversion of main magnetic contributor with the substitution of Ga for Co atom, which just means that the newly introduced atoms obey atomic preferential occupation rule. Based on the magnetic compensation effect, the new magnetic function materials with near zero molecular magnetic moment and higher $T_C$ can be explored experimentally. Unlike the case in Mn$_{3-x}$Co$_x$Ga alloys[52], the cubic structure can be still persisted over the whole substitution of Ga for Co atoms in Co$_{50-x}$Mn$_{25}$Ga$_{25+x}$. Thus, as a new physical phenomenon in Heusler alloys, the magnetic compensation behavior will provide new research field for a large class of Heusler compounds.

We next turn to discuss the half metallicity of $Co_{50-x}Mn_{25+x}Ga_{25}$. As shown in table II, we can find that the experimental and theoretical molecular magnetic moments all obey the Slater-Pauling rule[20, 33, 53, 54]. The magnetic moments of alloys show a monotonous decrease from 4 $\mu_B$ (for $Co_2MnGa$) to 2 $\mu_B$ (for $Mn_2CoGa$) [see Fig.5 (c)]. As mentioned in literature, the Slater-Pauling behavior is the important character of half metallic materials[20, 36, 53]. This means that the off-stoichiometric alloys in $Co_{50-x}Mn_{25+x}Ga_{25}$ are the possible half-metallic materials. And for $Co_{50-x}Mn_{25+x}Ga_{25}$ alloys, as shown in Fig.8, they are metallic for spin-up electrons and semionducting for spin-down electrons. Therefore, these ideal materials can be predicted as the $I_A$-type half metal[55]. Although there are only small density states near the $E_F$ in the spin-down direction and their spin polarizations are not 100%, these $Co_{50-x}Mn_{25+x}Ga_{25}$ alloys can be considered as nearly half metallictity with high polarization. So, these off-stoichiometric alloys between $Co_2MnGa$ and $Mn_2CoGa$ will provide new options for the research of compensated-ferrimagnetic half metallic materials[52]. The compensated-ferrimagnetic half metallic materials can be considered to have the potential advantage over the ferromagnetic half metals on their industrial application because they have no stray magnetic field and are much less affected by the external magnetic field[36, 56]. We can therefore conclude that these off-stoichiometric alloys reported in our work will proved more options for developing spintronic materials and attract particular attentions of researchers.

## V. CONCLUSIONS

In conclusion, we study the structure, magnetism, magnetic compensation behavior, exchange interaction and electronic structures of $Co_{50-x}Mn_{25}Ga_{25+x}$ and $Co_{50-x}Mn_{25+x}Ga_{25}$ ($x$=0-25) alloys. The main results are summarized as follows:

(1) In the composition intervals among Heusler alloy $Co_2MnGa$, $Mn_2CoGa$ and $CoMnGa_2$, the two series of off-stoichiometric alloys crystallize in single B.C.C structure with high atomic ordering. The occupation rule in principle dominates the atomic configuration in all samples, which causes the atomic preferential occupation not to be consistent with the chemical tendency in $Co_{50-x}Mn_{25}Ga_{25+x}$ system. The changes of lattice parameters indicate the dependence of them on atomic radius in the two series of samples. With increasing Ga content, the composition dependence of lattice parameter shows a kink point at the middle composition in $Co_{50-x}Mn_{25}Ga_{25+x}$ alloys, which can be attributed to the enhanced covalent effect.

(2) With the substitution of Mn for Co atoms in $Co_{50-x}Mn_{25+x}Ga_{25}$ alloys, the molecular magnetic moment monotonously decreases due to the local FIM structure established in the FM matrix. Owing to the role conversion of the main magnetic contributor, a collinear magnetic compensation behavior has been observed in $Co_{50-x}Mn_{25}Ga_{25+x}$ alloys. We further confirmed that the Heusler alloy $CoMnGa_2$ has a FIM structure in which the magnetic moment of Mn atom antiparallelly aligns with that of Co atom, showing a net molecular moment of 1.34 $\mu_B$.

(3) The measurement of Curie temperatures reveals the changes of the exchange interaction in the two series of off-stoichiometric alloys. The dominant behavior of FM or FIM structure has been observed in the different composition range of $Co_{50-x}Mn_{25+x}Ga_{25}$ alloys. In $Co_{50-x}Mn_{25}Ga_{25+x}$ alloys, the complicated magnetic competition between the FM and FIM structures leads to the Curie temperature dramatically decrease and shows a minimum value at $x$=20.

(4) The electronic structure calculations indicate that the $Co_{50-x}Mn_{25+x}Ga_{25}$ alloys tend to be in a coexistence state of the itinerant and localized magnetism. The broader DOS distribution for the *d*-electrons of Mn (*B*) atom results in an enhancement of hybridization between *d*-electrons of Mn ($A$) and Mn ($B$) atoms, which brings about an enhanced exchange interaction and the increased $T_C$ in $Co_{50-x}Mn_{25-x}Ga_{25}$ alloys. On the other hand, the covalent effect between Mn and Ga atoms enables the Mn ($A$) and Mn ($B$) atoms to have different atomic magnetic moments. Their antiparallel alignment results in the magnetic compensation behavior in the $Co_{50-x}Mn_{25}Ga_{25+x}$ alloys. The distinct hybridized peak near the Fermi level in the DOS of the Ga-rich alloys, especially of $CoMnGa_2$, implies an instable state, which may be responsible for the spin glass properties mentioned in the previous reports[21].

5) The compensated-ferrimagnetic alloys in $Co_{50-x}Mn_{25+x}Ga_{25}$ and $Co_{50-x}Mn_{25}Ga_{25+x}$ will prove more options for developing spintronic materials and attract particular attentions of the researchers.

Acknowledgement

This work is supported by the National Natural Science Foundation of China in Grant No. 51021061, 51071172 and 11174352 and National Basic Research Program of China (973 Program, 2010CB833102 and 2012CB619405).

Figure captions:

FIG.1. (Color online) The crystal structure of Full-Heusler compounds $X_2YZ$ with four interpenetrating f.c.c sublattices of $A$, $B$, $C$ and $D$.

FIG.2. (color online) DSC curves of $Co_{50-x}Mn_{25}Ga_{25+x}$ (a) and $Co_{50-x}Mn_{25+x}Ga_{25}$ (b) alloys. The $T_t$ and $T_m$ represent the order-disorder transition temperature, i.e., from $L2_1$ to $B2$ or $Hg_2CuTi$-type to $B2$ structure, and melting temperature, respectively.

FIG.3. (Color online) XRD patterns of the $Co_{50-x}Mn_{25}Ga_{25+x}$ (a) and $Co_{50-x}Mn_{25+x}Ga_{25}$ (b) alloys ($x$=0, 7.5, 12, 17.5, 25). The left column of each panel shows the superlattice peaks of (111) and (200) and the right column is for the main peaks. Insets on the top right corner show the composition dependence of the degree of ordering S. The open triangle and square represent the data taken from literature[21]and [34], respectively.

FIG.4. (Color online) The composition dependence of the lattice parameters of the $Co_{50-x}Mn_{25}Ga_{25+x}$ and $Co_{50-x}Mn_{25+x}Ga_{25}$ ($x$=0-25) alloys. The arrow indicates the kink behavior. The inset is the structure of $CoMnGa_2$ Heusler alloy.

FIG.5. (Color online) (a) and (b) show the representative magnetization curves of $Co_{50-x}Mn_{25}Ga_{25+x}$ (left column for $x \leqq 15$ and right column for $x$>15) and $Co_{50-x}Mn_{25+x}Ga_{25}$ ($x$=0-25) alloys measured at 5 K, respectively. (c) The composition dependence of molecular magnetic moments, measured (solid triangle) and calculated (open triangle), of $Co_{50-x}Mn_{25}Ga_{25+x}$

and $Co_{50-x}Mn_{25+x}Ga_{25}$ (x=0-25) alloys.

FIG.6. (Color online) The schematic diagram of the magnetic structure changes for $Co_{50-x}Mn_{25+x}Ga_{25}$ (x=0-25) and $Co_{50-x}Mn_{25}Ga_{25+x}$ (x=0-25), respectively.

FIG.7 (Color online) The composition dependence of Curie temperature of $Co_{50-x}Mn_{25+x}Ga_{25}$ (a) and $Co_{50-x}Mn_{25}Ga_{25+x}$ (b) (x=0-25) alloys. The inset of (a) shows the typical temperature dependence of magnetization curves of $Co_{50-x}Mn_{25+x}Ga_{25}$ (x=2.5, 7.5, 12, 22.5) alloys measured under 100 Oe. The inset of (b) on the bottom left corner shows the typical temperature dependence of magnetization curves of $Co_{50-x}Mn_{25}Ga_{25+x}$ (x=0, 5, 7.5, 10) measured under 100 Oe. While the inset of (b) on the top right corner shows the typical temperature dependence of magnetization curves of $Co_{50-x}Mn_{25}Ga_{25+x}$ (x=20, 25) under zero field cooling (solid circles) and field cooing (open circles), respectively. The Curie temperature was determined from the minimum value of dM(T)/dT curves.

FIG.8. (Color online) Calculated spin-projected DOS plots for the $Co_{50-x}Mn_{25+x}Ga_{25}$ alloys. The total DOS (a), the partial DOS for *d*-electrons of Mn (*A*), Mn (*B*), Co (*A*) ,Co (*C*) atoms and for *p*-electrons of Ga (*D*) atoms are illustrated in (a)～(f), respectively. The upper halves of each panel display the spin-up states.

FIG.9. (Color online) Calculated spin-projected DOS plots for the $Co_{50-x}Mn_{25}Ga_{25+x}$ alloys. The total DOS (a), the partial DOS for *d*-electrons of Mn (*A*), Mn (*B*), Co (*A*) ,Co (*C*) atoms and for *p*-electrons of Ga (*B*) and Ga (*D*) atoms are illustrated in (a)～(g), respectively. The upper halves of each panel display the spin-up states.

Table captions:

TABLE I. The lattice parameters of $Co_{50-x}Mn_{25}Ga_{25+x}$ and $Co_{50-x}Mn_{25+x}Ga_{25}$ ($x$=0-25) alloys with various compositions. Here a and a* denote the lattice parameters of $Co_{50-x}Mn_{25}Ga_{25+x}$ and $Co_{50-x}Mn_{25+x}Ga_{25}$ samples, respectively.

| $x$ | a(nm) | a*(nm) |
|---|---|---|
| **0** | 0.5764 | 0.5764 |
| **2.5** | 0.5778 | 0.5771 |
| **5** | 0.5790 | 0.5782 |
| **7.5** | 0.5799 | 0.5791 |
| **10** | 0.5809 | 0.5805 |
| **12** | 0.5820 | 0.5812 |
| **13** | 0.5823 | 0.5817 |
| **15** | 0.5827 | 0.5826 |
| **17.5** | 0.5831 | 0.5835 |
| **20** | 0.5835 | 0.5844 |
| **22.5** | 0.5839 | 0.5853 |
| **25** | 0.5846 | 0.5865 |

TABLE II. The calculated molecular magnetic moments ($M_{calc.}$) and measured molecular magnetic moments ($M_{exp.}$) (in $\mu_B$/f.u.), as well as the calculated atomic magnetic moments (in $\mu_B$/atom) for $Co_{50-x}Mn_{25+x}Ga_{25}$ alloys with various compositions.

| $x$ | Co($A$) | Mn ($A$) | Mn ($B$) | Co ($C$) | Ga ($D$) | $M_{cal.}$ | $M_{exp.}$ |
|---|---|---|---|---|---|---|---|
| 0 | 0.70 | -- | 2.89 | 0.70 | -0.12 | 4.19 | 4.14 |
| 2.5 | 0.74 | -1.94 | 2.87 | 0.72 | -0.11 | 3.95 | 3.87 |
| 5 | 0.77 | -1.89 | 2.86 | 0.73 | -0.10 | 3.73 | 3.7 |
| 7.5 | 0.81 | -1.83 | 2.84 | 0.75 | -0.09 | 3.51 | 3.55 |
| 10 | 0.84 | -1.79 | 2.84 | 0.76 | -0.09 | 3.30 | 3.39 |
| 12 | 0.86 | -1.74 | 2.82 | 0.77 | -0.08 | 3.13 | 3.15 |
| 13 | 0.87 | -1.74 | 2.83 | 0.78 | -0.08 | 3.04 | 2.97 |
| 15 | 0.89 | -1.72 | 2.83 | 0.79 | -0.07 | 2.87 | 2.89 |
| 17.5 | 0.92 | -1.73 | 2.86 | 0.80 | -0.07 | 2.66 | 2.71 |
| 20 | 0.95 | -1.71 | 2.87 | 0.82 | -0.06 | 2.45 | 2.44 |
| 22.5 | 0.97 | -1.69 | 2.88 | 0.84 | -0.06 | 2.24 | 2.34 |
| 25 | -- | -1.67 | 2.88 | 0.88 | -0.05 | 2.03 | 2.10 |

TABLE III. The calculated molecular magnetic moments ($M_{calc.}$) and measured molecular magnetic moments ($M_{exp.}$) (in $\mu_B$/f.u.), as well as the calculated atomic magnetic moments (in $\mu_B$/atom) for $Co_{50-x}Mn_{25}Ga_{25+x}$ alloys with various compositions.

| $x$ | Co (*A*) | Mn (*A*) | Mn (*B*) | Ga (*B*) | Co (*C*) | Ga (*D*) | $M_{calc.}$ | $M_{expr.}$ |
|---|---|---|---|---|---|---|---|---|
| 0 | 0.70 |  | 2.89 |  | 0.70 | -0.12 | 4.19 | 4.14 |
| 2.5 | 0.66 | -1.91 | 2.93 | -0.09 | 0.64 | -0.10 | 3.58 | 3.55 |
| 5 | 0.63 | -1.82 | 2.96 | -0.07 | 0.58 | -0.09 | 2.99 | 2.97 |
| 7.5 | 0.61 | -1.71 | 2.94 | -0.05 | 0.52 | -0.06 | 2.42 | 2.36 |
| 10 | 0.58 | -1.65 | 2.94 | -0.03 | 0.45 | -0.05 | 1.84 | 1.93 |
| 13 | 0.53 | -1.57 | 2.91 | -0.02 | 0.37 | -0.03 | 1.16 | 1.25 |
| 15 | 0.49 | -1.55 | 2.89 | -0.01 | 0.31 | -0.02 | 0.72 | 0.61 |
| 17.5 | 0.48 | -1.57 | 2.89 | -0.00 | 0.26 | -0.01 | 0.16 | 0.17 |
| 20 | -0.47 | 1.60 | -2.86 | -0.00 | -0.22 | -0.00 | 0.39 | 0.26 |
| 22.5 | -0.47 | 1.67 | -2.86 | -0.01 | -0.19 | -0.00 | 0.97 | 0.70 |
| 25 | -- | 1.78 | -- | -0.01 | -0.19 | -0.01 | 1.57 | 1.34 |

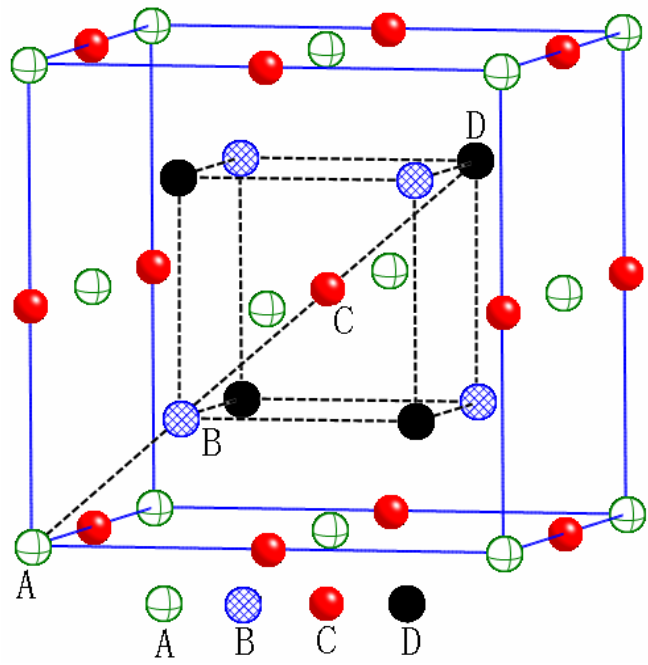

Figure 1, G. J. Li et al., for PRB

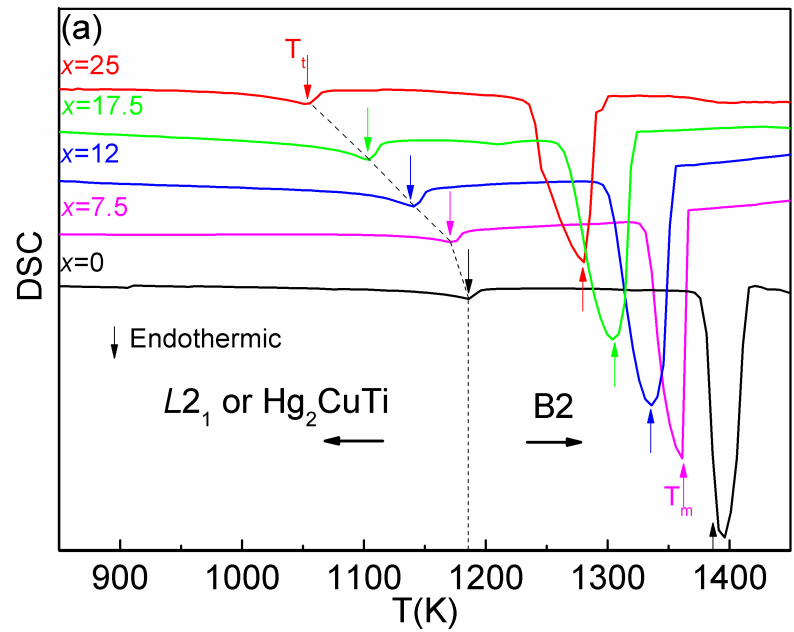

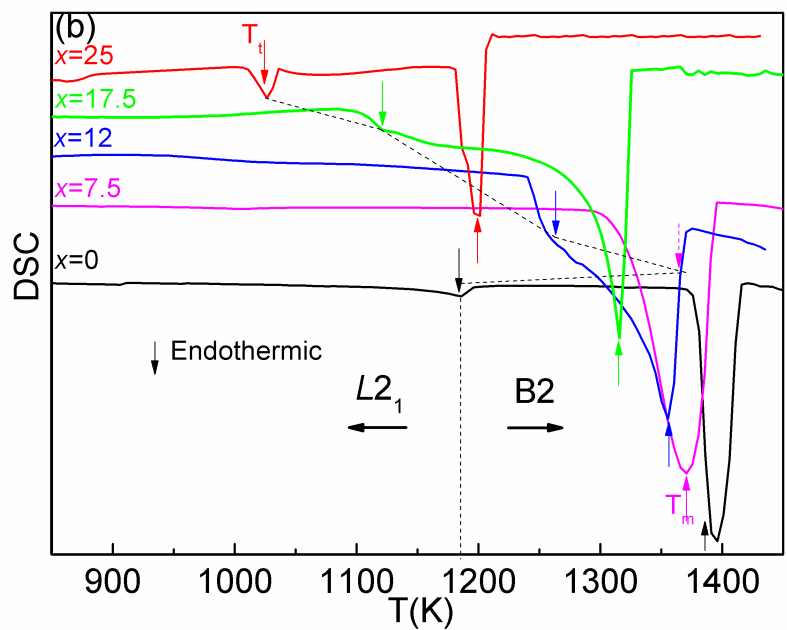

Figure 2, G. J. Li et al., for PRB

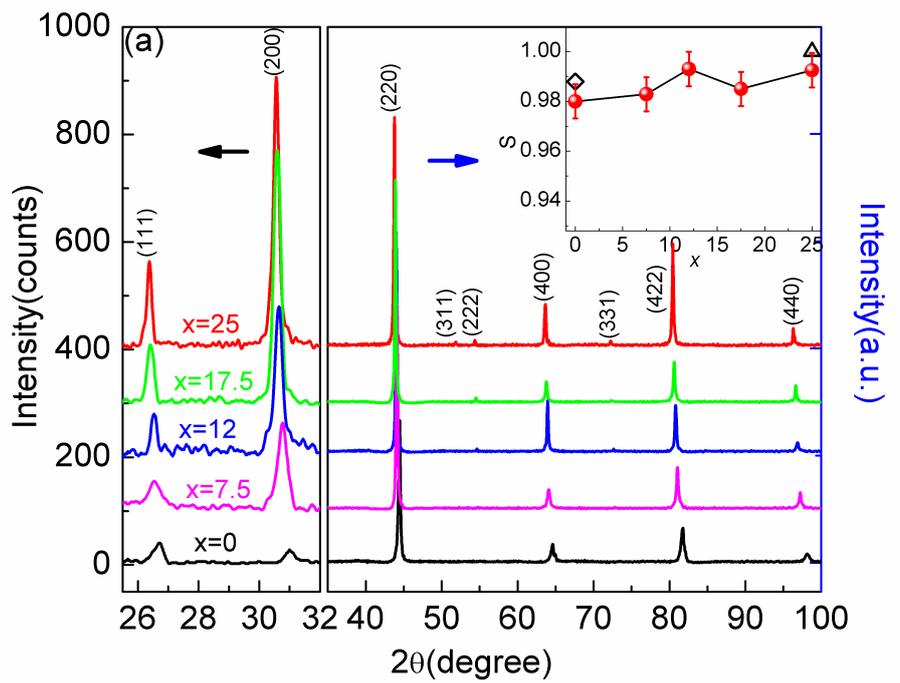

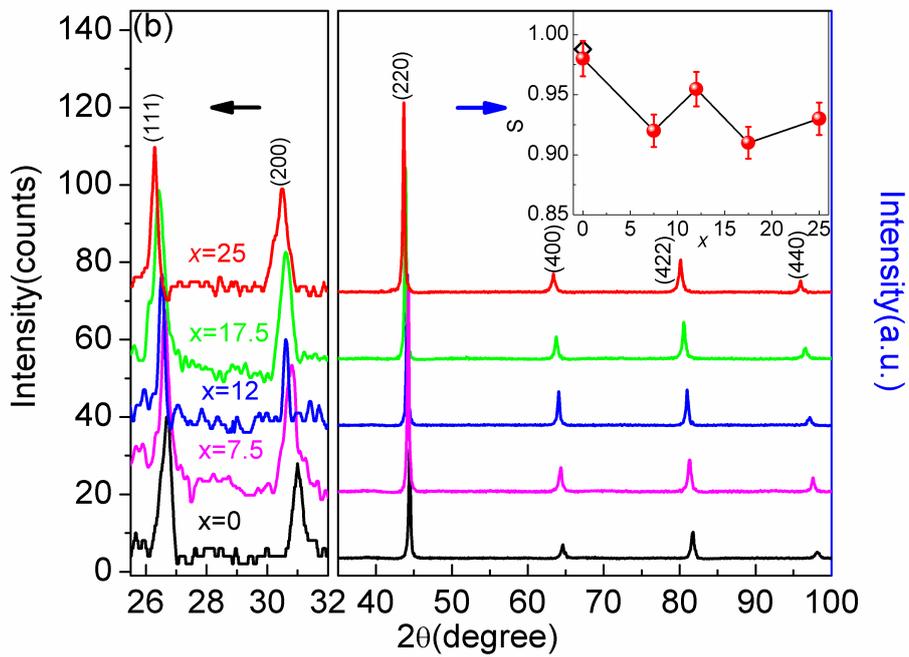

Figure 3, G. J. Li et al., for PRB

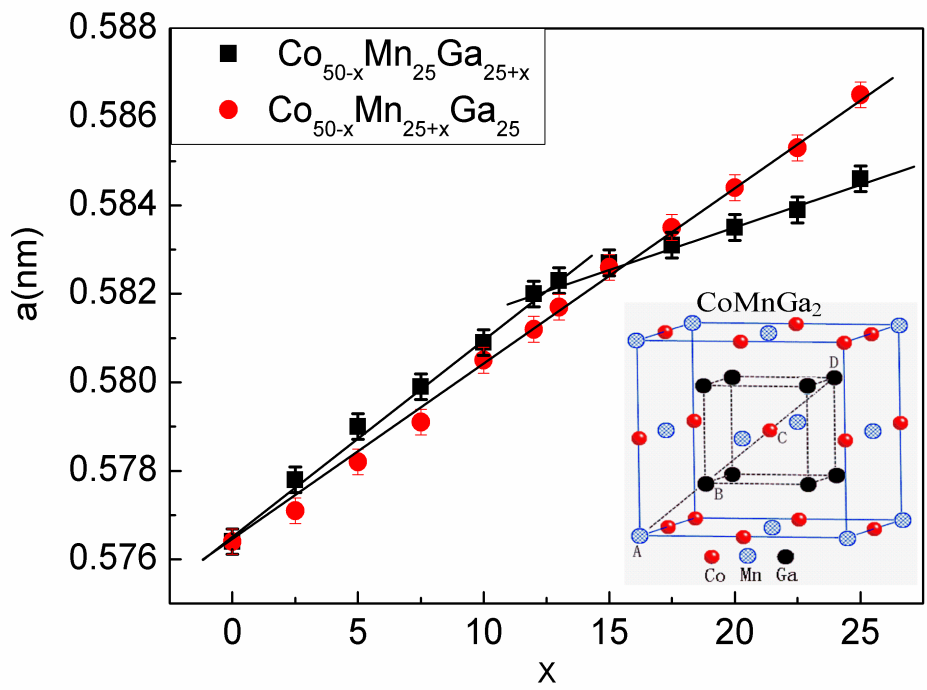

Figure 4, G. J. Li et al., for PRB

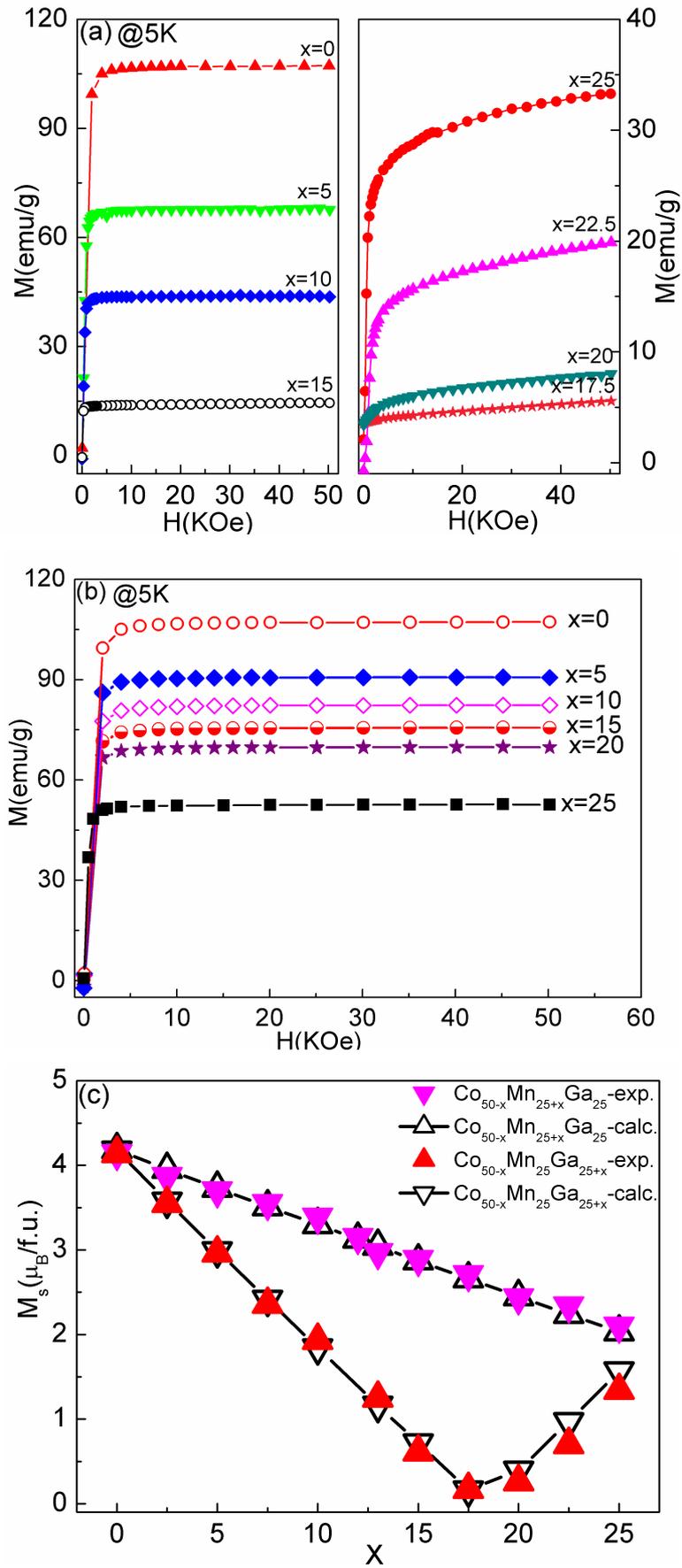

Figure 5, G. J. Li et al., for PRB

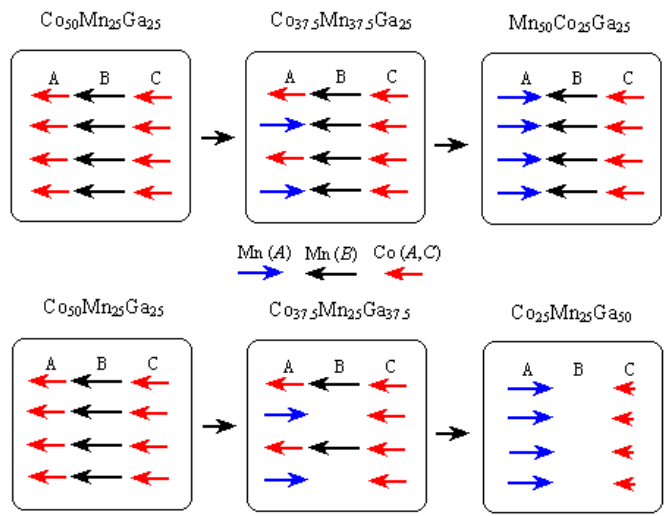

Figure 6, G. J. Li et al., for PRB

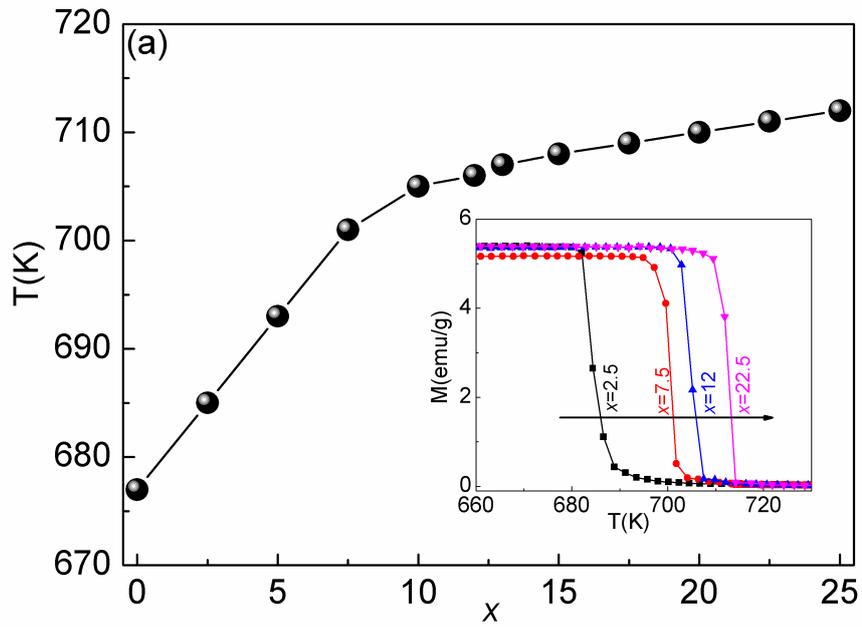

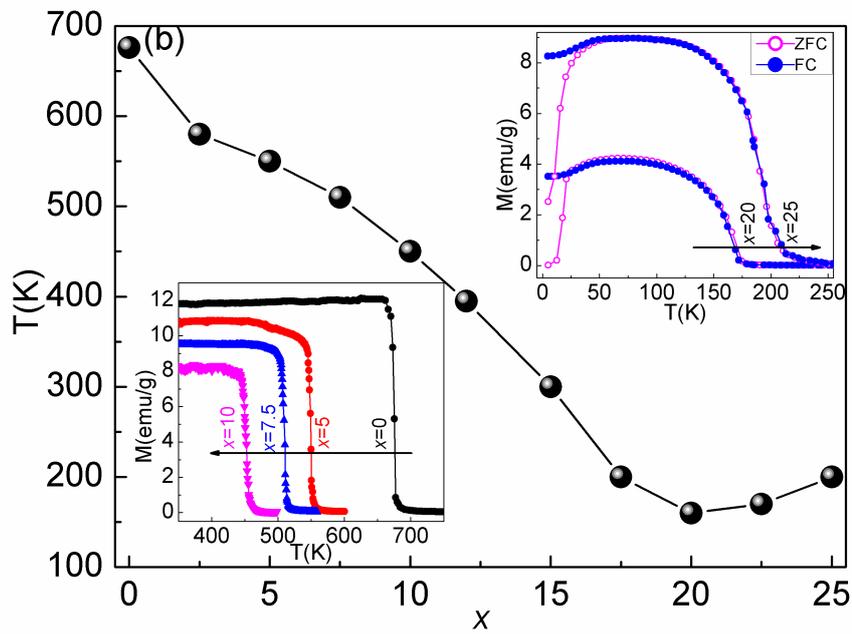

Figure 7, G. J. Li et al., for PRB

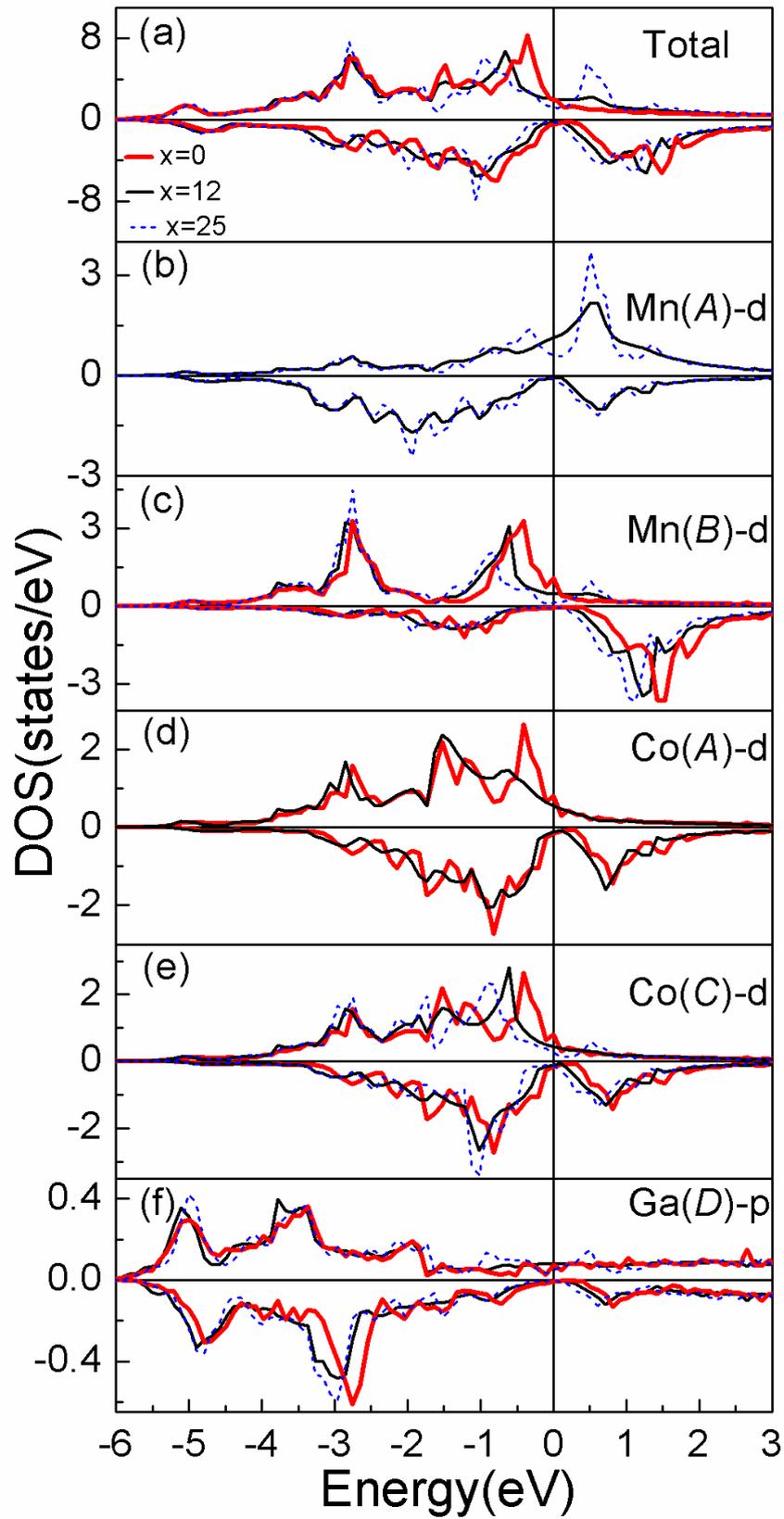

Figure 8, G. J. Li et al., for PRB

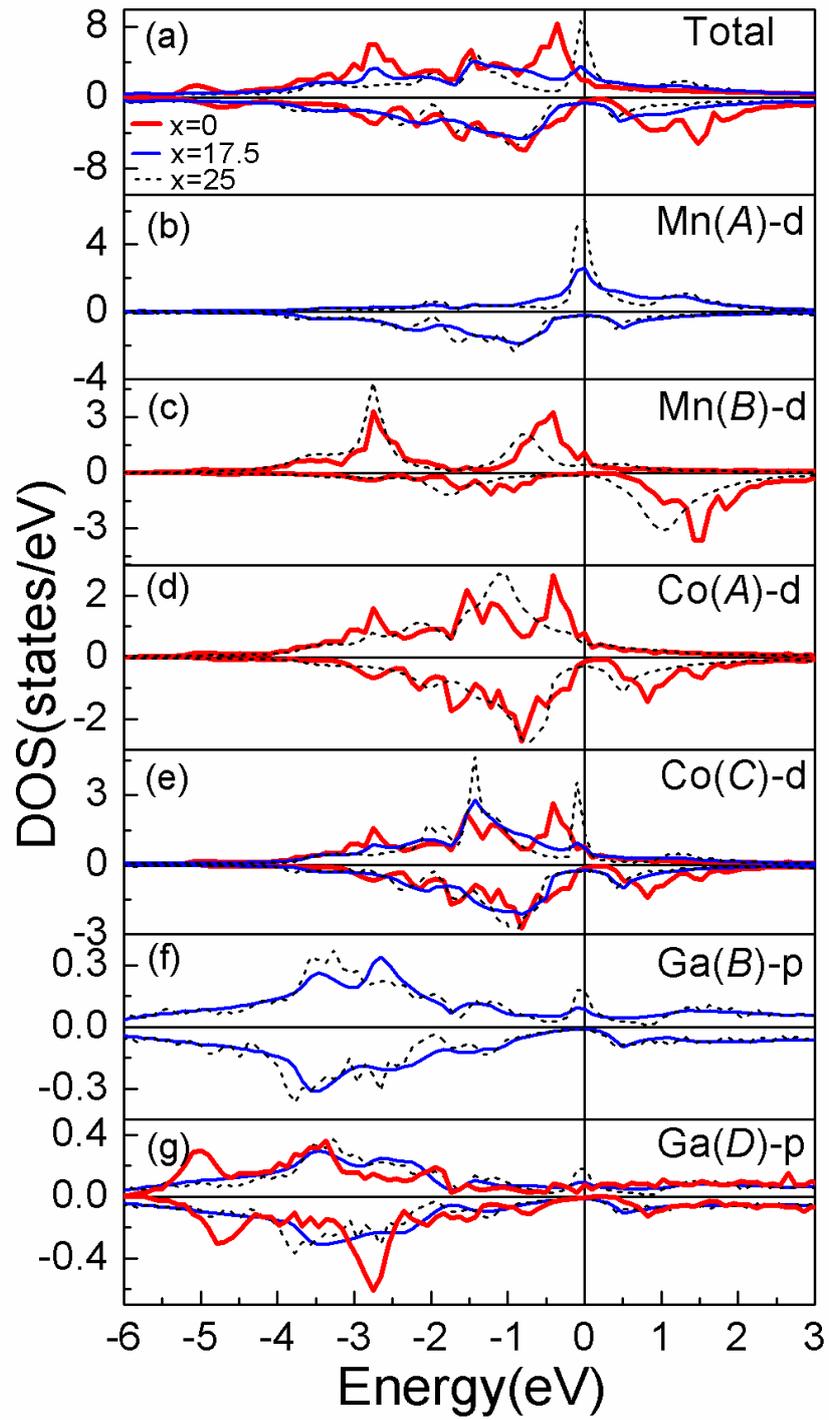

Figure 9, G. J. Li et al., for PRB

# Supplementary information:

1) The simulated and experimental XRD patterns

In Heusler alloy, the four interpenetrating f.c.c. sublattices $A$, $B$, $C$ and $D$ gives rise to non-zero Bragg reflections only when the Miller indices of the scattering planes are either all even, **or** all odd. The even planes may be sub-divided into two groups, those for which $(h+k+l)/2$ is odd, and those for which $(h+k+l)/2$ is even. The intensities of these reflections are determined by the squares of the structure factors, $F$, which for the three types of reflection may be shown to be

$$F(111) = 4\left|(f_A - f_C)^2 + (f_B - f_D)^2\right|^{1/2}$$

$$F(200) = 4\left|f_A - f_B + f_C - f_D\right|$$

$$F(220) = 4\left|f_A + f_B + f_C + f_D\right|$$

where $f_A$, $f_B$, $f_C$ and $f_D$ are the average scattering factors of the $A,B,C$ and $D$ sites respectively.

$F(111)$ and $F(200)$ contain difference terms and hence correspond to the order-dependent superlattice reflections which would be absent if all the average scattering factors were equal. $F(220)$ is the sum of the scattering factors and is independent of order. Such reflections are known as the principal reflections.

A. In $Co_{50-x}Mn_{25+x}Ga_{25}$ alloys

(i) The $Co_2MnGa$ alloy with $L2_1$ structure

$$F(111) = 4\left|(f_A - f_C)^2 + (f_B - f_D)^2\right|^{1/2} = 4\left|(f_{Co} - f_{Co})^2 + (f_{Mn} - f_{Ga})^2\right|^{1/2} = 4\left|f_{Mn} - f_{Ga}\right|$$

$$F(200) = 4\left|f_A - f_B + f_C - f_D\right| = 4\left|2f_{Co} - f_{Mn} - f_{Ga}\right|$$

One can find that the intensity of (111) is light more than that of (200). As shown in the Fig.S1(c) and (f).

(ii) The $Co_{1.5}Mn_{1.5}Ga$ alloy

For $Co_{1.5}Mn_{1.5}Ga$ alloy, the intensity of (111) and (200) superlattice peaks can be

described as below:

$$F(111) = 4\left|(f_A - f_C)^2 + (f_B - f_D)^2\right|^{1/2} = 4\left|(\frac{1}{2}f_{Co} + \frac{1}{2}f_{Mn} - f_{Co})^2 + (f_{Mn} - f_{Ga})^2\right|^{1/2}$$

$$= 4\left|\frac{1}{4}(f_{Mn} - f_{Co})^2 + (f_{Mn} - f_{Ga})^2\right|^{1/2}$$

$$F(200) = 4\left|f_A - f_B + f_C - f_D\right| = 4\left|\frac{1}{2}f_{Co} + \frac{1}{2}f_{Mn} - f_{Mn} + f_{Co} - f_{Ga}\right|$$

$$= 4\left|\frac{1}{2}(f_{Co} - f_{Mn}) + f_{Co} - f_{Ga}\right|$$

The intensity of (111) is more than that of (200), which is basically consistent with the simulated results.

If the $A$ and $C$ sites were equally occupied by the Mn and Co atoms, then the intensity of (111) and (200) can be rewritten as

$$F(111) = 4\left|(f_A - f_C)^2 + (f_B - f_D)^2\right|^{1/2} = 4\left|(f_{Mn} - f_{Ga})^2\right|^{1/2} \quad \text{and}$$

$$F(200) = 4\left|f_A - f_B + f_C - f_D\right| = 4\left|\frac{1}{2}(f_{Co} - f_{Mn}) + f_{Co} - f_{Ga}\right|$$

Obviously the intensity of (111) is lowered while that of (200) is hardly changed. Maybe due to the partial disorder occupation between $A$ and $C$, the intensity ratio between (111) and (200) become lower as shown in Fig.S1(b) and (e). The influence of disorder occupation on the magnetic properties will be discussed in the following section.

(iii) The Mn$_2$CoGa alloy with Hg$_2$CuTi-type structure

The intensity of (111) and (200) superlattice peaks are as following

$$F(111) = 4\left|(f_A - f_C)^2 + (f_B - f_D)^2\right|^{1/2} = 4\left|(f_{Mn} - f_{Co})^2 + (f_{Mn} - f_{Ga})^2\right|^{1/2}$$

$$F(200) = 4\left|f_A - f_B + f_C - f_D\right| = 4\left|f_{Mn} - f_{Mn} + f_{Co} - f_{Ga}\right| = 4\left|f_{Co} - f_{Ga}\right|$$

So $F(111)^2 - F(200)^2 = 32(f_{Mn} - f_{Ga})(f_{Mn} - f_{Co}) > 0$, which indicate the intensity of (111) is higher than that of (200).

If the Co and Mn equally occupy $A$ and $C$ site in Mn$_2$CoGa alloy, the intensity of (111)

and (200) can be

$$F(111) = 4\left|(f_A - f_C)^2 + (f_B - f_D)^2\right|^{1/2} = 4\left|f_{Mn} - f_{Ga}\right|$$

$$F(200) = 4\left|f_A - f_B + f_C - f_D\right| = 4\left|f_{Mn} - f_{Mn} + f_{Co} - f_{Ga}\right| = 4\left|f_{Co} - f_{Ga}\right|$$

$F(111) - F(200) > 0$. Although the intensity of (111) is still more than that of (200), the relative intensity ratio of the superlattice peaks (111) and (200) become lower, which is closely related with this disorder occupation.

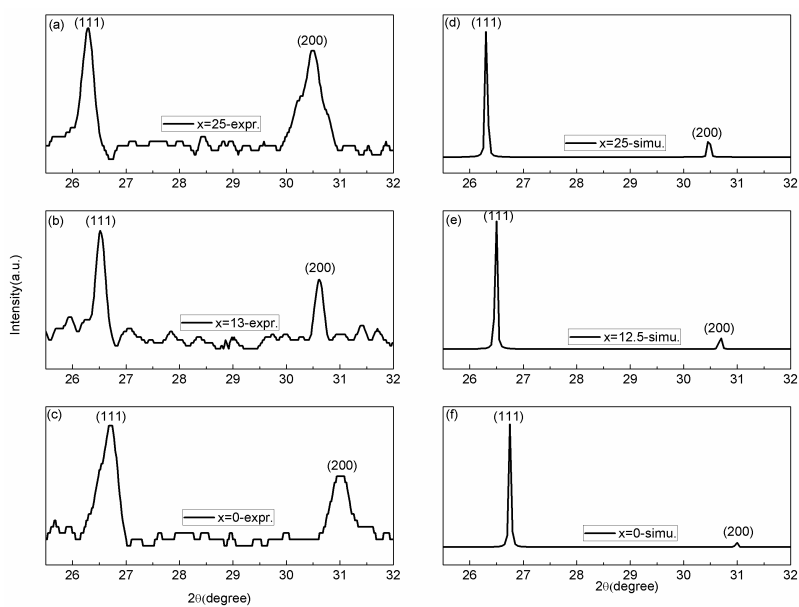

Fig. S1 the experimental (a-c) and simulated (d-f) XRD results of $Co_{50-x}Mn_{25+x}Ga_{25}$ alloys.

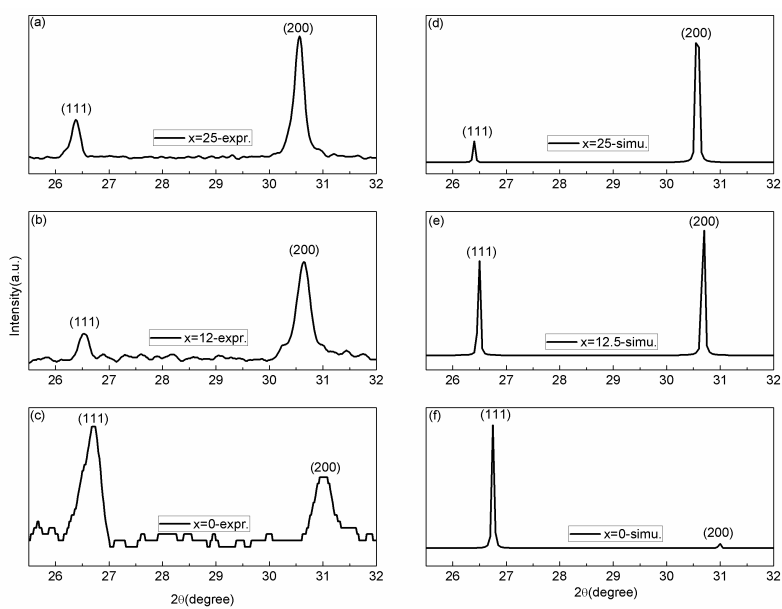

Fig. S2 the experimental (a-c) and simulated (d-f) XRD results of $Co_{50-x}Mn_{25}Ga_{25+x}$

alloys.

B. In $Co_{50-x}Mn_{25}Ga_{25+x}$ alloys

(i) In $CoMnGa_2$ alloy with $L2_1$ structure

$$F(111) = 4\left|(f_A - f_C)^2 + (f_B - f_D)^2\right|^{1/2} = 4\left|(f_{Ga} - f_{Ga})^2 + (f_{Co} - f_{Mn})^2\right|^{1/2} = 4\left|f_{Co} - f_{Mn}\right|$$

$$F(200) = 4\left|2f_{Ga} - (f_{Co} + f_{Mn})\right|$$

Obviously the intensity of the $F(200)$ is much more than that of $F(111)$.

Importantly, in our work, the experiment and simulation results are well consistent with each other, which indicate $CoMnGa_2$ alloy crystallized in high-ordered $L2_1$ structure.

In the $CoMnGa_2$, if the Co and Mn equally occupy $A$ and $C$ sites, the intensity of (111) reflection peak can be described as following:

$$F(111) = 4\left|f_{Co} - f_{Mn}\right| = 4\left|(\tfrac{1}{2}f_{Co} + \tfrac{1}{2}f_{Mn}) - (\tfrac{1}{2}f_{Mn} + \tfrac{1}{2}f_{Co})\right|$$

So the $F(111)$ intensity is closed to zero.

While the intensity of (200) reflection peak

$$F(200) = 4\left|2f_{Ga} - (f_{Co} + f_{Mn})\right| = 4\left|2f_{Ga} - ((\tfrac{1}{2}f_{Co} + \tfrac{1}{2}f_{Mn} + \tfrac{1}{2}f_{Mn} + \tfrac{1}{2}f_{Co}))\right|$$ can be

considered as unchanged. These two analyses are all well consistent with the experimental and calculated result, strongly suggesting the $CoMnGa_2$ alloy forms in high-ordered $L2_1$ structure.

(ii) In the case of $Co_{37.5}Mn_{25}Ga_{37.5}$ alloy

$$F(111) = 4\left|(f_A - f_C)^2 + (f_B - f_D)^2\right|^{1/2} = 4\left|(\tfrac{1}{2}f_{Co} + \tfrac{1}{2}f_{Mn} - f_{Co})^2 + (\tfrac{1}{2}f_{Mn} + \tfrac{1}{2}f_{Ga} - f_{Ga})^2\right|^{1/2}$$

$$= 4\left|\tfrac{1}{4}(f_{Co} - f_{Mn})^2 + \tfrac{1}{4}(f_{Mn} - f_{Ga})^2\right|^{1/2}$$

$$F(200) = 4\left|f_A - f_B + f_C - f_D\right| = 4\left|\tfrac{3}{2}(f_{Co} - f_{Ga})\right|$$

Obviously, the intensity of $F(200)$ is more than that of $F(111)$, which is consistent with the experimental result, as shown in Fig.S2(b) and (e)

Summary, in the two serials of alloys, the experimental XRD results are all well consistent with the simulated ones, which strongly suggests all the samples state in high-ordered structure or the high-ordered structure dominate all the samples.

2) Influence of heat-treatment temperature on the ordering parameter S

In order to obtain heat-treatment temperature dependence of ordering parameter S, the heat treatments under different ordering temperature were carried out. Here, we took $Co_{50-x}Mn_{25+x}Ga_{25}$ alloys for example. And then the ordering parameters S were calculated based on the experimental and simulated XRD results. These results were list in Table SI. From these data, we can find that the values of S for the samples with ordering temperature of 973 K are all higher than those with the ordering temperature of 673 K, even if one week's ordering treatment. This strongly suggests the ordering temperature just a bit lower than transition temperature should be appropriate.

TABLE.SI the ordering parameters S of $Co_{50-x}Mn_{25+x}Ga_{25}$ alloys under different heat-treatment temperatures

| $x$ | 0 | 7.5 | 12 | 17.5 | 25 |
|---|---|---|---|---|---|
| S1 | 0.988 | 0.92 | 0.955 | 0.91 | 0.930 |
| S2 | 0.823 | 0.84 | 0.847 | 0.73 | 0.787 |

S1 denotes ordering parameter of samples from the heat-treatment at 1073 K for two days, at 973 K for one day.
S2 denotes ordering parameter of samples from the heat-treatment at heat-treatment at 1073 K for two days, at 673 K for one week.